\documentclass{jfm}

\renewcommand{\absfooterflag}{}
\renewcommand{\pagelimitfooter}{}
\usepackage{graphicx}
\usepackage{newtxtext}
\usepackage{newtxmath}
\usepackage{natbib}
\usepackage{hyperref}
\usepackage{import}

\usepackage{caption}
\usepackage{subcaption}
\usepackage{cleveref}
\usepackage{multicol}
\usepackage{bm}
\usepackage{xcolor}

\newcommand{\Reynolds}{\mathrm{Re}}
\newcommand{\Prandtl}{\mathrm{Pr}}
\newcommand{\Ri}{\mathrm{Ri}}
\newcommand{\Rig}{\mathrm{Ri}_\mathrm{g}}

\newcommand{\diff}{\ensuremath{\; d}}

\hypersetup{
    colorlinks = true,
    urlcolor   = blue,
    citecolor  = black,
}

\newcommand{\RomanNumeralCaps}[1]

\title{Early onset of secondary shear instability in Kelvin–Helmholtz braids at high Reynolds number}

\author{Emma R. Bouckley\aff{1}
  \corresp{\email{eb775@cam.ac.uk}},
  Sam F. Lewin \aff{2}
 \and Adrien Lefauve\aff{3,4}}

\begin{document}
\maketitle

\begin{abstract}
We study the onset of two-dimensional secondary shear instability (SSI) in the braid regions connecting primary Kelvin-Helmholtz billows in stratified shear flows. While  strain induced by the billows stabilises the braids, it also compresses their tilted isopycnals, enhancing baroclinic shear that enables rapid perturbation growth.
By modifying the classical analysis of Corcos \& Sherman (\textit{J. Fluid Mech.} \textbf{73}, 241-264, 1976) in braid-aligned coordinates and adding an additional stability criterion based on the ratio of strain rate to shear,
we develop an inviscid, time-dependent model for the braid and the onset of SSI.  
We show that the criterion for instability can be achieved significantly earlier than the saturation of the primary billow at sufficiently high initial Richardson number $\Ri$, as increased stratification slows billow growth while accelerating baroclinic shear production in the braid. Two-dimensional direct numerical simulations up to  Reynolds numbers $\Reynolds=10^7$ quantify the role of viscosity. At  high $\Reynolds$, we find that SSI indeed develops early in the braid, as predicted by the inviscid model, while the primary billow is still growing and before viscosity slows braid thinning. These results provide a mechanistic explanation for field observations of braid-dominated mixing
and suggest that, at geophysically relevant $\Ri$ and $\Reynolds$,  SSI can control the three-dimensional turbulent transition and ensuing diapycnal mixing by preceding and pre-empting both vortex pairing instabilities and secondary convective instabilities in the billow core.
\end{abstract}



\section{Introduction}

The transition to turbulence from Kelvin–Helmholtz (KH) instability in stably stratified, parallel shear flows is a long-standing open problem in fluid dynamics with broad geophysical and astrophysical relevance. Classically, this transition is assumed to occur after normal mode perturbations grow to finite amplitude, producing quasi two-dimensional vortex structures (referred to as billows) that subsequently collapse, resulting in the rapid onset of three-dimensional motions. Importantly, the properties of subsequent turbulence and mixing are sensitive to the details of the transition \citep{caulfield_layering_2021, mashayek_shear-induced_2013}.

At high Reynolds number, the finite amplitude billow structure is subject to a wide range of secondary instabilities \citep{mashayek_zoo_2012-1,mashayek_zoo_2012} that may be grouped broadly into three classes. The first class consists of vortex pairing instabilities (e.g. \citealp{klaassen_role_1989}) occurring at subharmonics of the most unstable primary streamwise wavenumber. The second class extracts their energy from unstable density gradients in the billow core and are referred to as secondary convective instabilities (e.g. \citealp{klaassen_onset_1985}). The third class instead extracts their energy from the narrow, inclined strips of vorticity connecting adjacent billows, referred to as braids \citep{corcos_vorticity_1976,staquet_two-dimensional_1995,smyth_secondary_2003}. Though a range of braid instabilities are possible \citep{mashayek_zoo_2012-1,mashayek_zoo_2012}, this study focuses specifically on secondary shear instability (SSI), which produces trains of vortices that are advected along the braid and towards the cores. Turbulent transition is governed by a competition between the growth rates of these instabilities, which evolve significantly as the primary billow grows and saturates  in amplitude.


In the ocean, KH instability is considered a primary pathway between submesoscales and stratified turbulence \citep{smyth_ocean_2012}, with coherent billow structures observed from the deep ocean to the thermocline, continental shelves and estuaries. In some cases, detailed acoustic, velocity and density observations \citep{geyer_mixing_2010,vladoiu_finescale_2025,lefauve_structure_2026b} show intense turbulent mixing concentrated in the braids, with large overturns rarely present. A compilation of observations from the past four decades (Appendix A of \cite{lefauve_structure_2026b}) suggests that such braid-dominated mixing is widespread across stratification types (temperature, salinity, and turbidity interfaces) and shear-layer thicknesses spanning $\approx 0.5$–$200$ m, corresponding to Reynolds numbers $\approx 10^5$–$10^8$. Providing a mechanistic and predictive basis for the hypothesis of \cite{geyer_mixing_2010} that mixing is braid-dominated at these Reynolds numbers, the present study clarifies why and when SSI is expected to control the transition to three-dimensional turbulence, thereby shaping subsequent mixing across a range of ocean flows.


Seminal early work on SSI by \cite{corcos_vorticity_1976} analysed the local flow in a tilted, braid-aligned coordinate system, assuming the resulting instabilities to be  of the classical inflectional shear type. Subsequently, \cite{staquet_two-dimensional_1995} drew on the stability of a strip of vorticity under strain \citep{dritschel_stability_1991} to propose a modified braid stability criterion based on the ratio of strain to shear, later adopted by \citet{smyth_secondary_2003} and \citep{mashayek_zoo_2012-1,mashayek_zoo_2012}. Importantly, these later studies assumed either a steady or frozen-in-time background flow following saturation of the primary billow. Here, we relax this assumption and explicitly model the time-evolving steepening of the braid. This distinction is essential because, as we will show, at geophysical Reynolds numbers SSI can develop well before billow saturation and must therefore be predicted from the evolving braid.

To this end, we introduce an inviscid model for the time evolution and stability of the braid in \cref{modelsetup}, followed by two-dimensional numerical simulations in \cref{numsim}. In \cref{results}, we present results for the onset of SSI across a range of Richardson and Reynolds numbers and examine the agreement with the inviscid model. Finally, we conclude in \cref{discussion}.

\section{Theory}\label{modelsetup}

\subsection{Governing equations and parameters}

We consider a two-dimensional, stably stratified, vertically sheared layer evolving in time $t$ with streamwise coordinate $x$ and vertical coordinate $z$. 
The flow is incompressible, with small density variations under the Boussinesq approximation. The dimensionless governing equations and initial conditions are
\begin{align}
    \nabla\cdot \bm u =0,\qquad \frac{D  \bm u}{D t} = -\nabla p+ \Ri  \,b\, \bm{ \hat {z}} + \frac{1}{\Reynolds} \nabla^2 \bm u,
    \qquad\frac{D b}{D t} = \frac{1}{ \Reynolds \,\Prandtl} \nabla ^2 b,\label{govnd}
    \\ \bm u(z,\; t=0)=  \tanh(z)\,\bm{\hat x},
    \qquad b(z,\;t=0) = \tanh(z).\label{ICnd}
\end{align}
All lengths are scaled by the shear layer half-depth $d$, whilst velocities and buoyancy are scaled with half the dimensional velocity and buoyancy differences across the shear layer $u_0$ and $b_0$. 
\Cref{govnd,ICnd} include three dimensionless numbers, the Reynolds number $\Reynolds$, the Richardson number $\Ri$, and  the Prandtl number $\Prandtl$:
\begin{equation}
    \Reynolds = \frac{u_0 \,d}{\nu}, \qquad \Ri =\frac{ b_0\,d}{u_0^2},  \qquad \Prandtl=\frac{\nu}{\kappa}=1,
\end{equation}
where $\nu$ and $\kappa$ are the kinematic viscosity and buoyancy diffusivity, respectively. 

\subsection{Inviscid model of braid evolution}\label{theoreticalbraidanalysis}

Following \cite{corcos_vorticity_1976}  (hereafter CS76), we begin by defining a braid-aligned coordinate system tilted at an average angle $\phi(t)$ to the horizontal. The along-braid variation of the angle is insignificant (as we will see). The along-braid coordinate $x'$ and cross-braid  coordinate $z'$ are defined by
\begin{align}
    x' = x\cos(\phi)+z\sin(\phi),
    \qquad z' = -x\sin(\phi)+z\cos(\phi),
\end{align}
where the inflection point of the braid is located at the origin $(x,z)=(x',z')=(0,0)$. Neglecting the time-dependent rotation of the coordinate system, the governing equations (\ref{govnd}) in braid coordinates can be written as
\begin{align}
    \frac{\partial b}{\partial t} &= -u'\frac{\partial b}{\partial x'}-w'\frac{\partial b}{\partial z'} + \frac{1}{\Reynolds \,\Prandtl}\nabla'^2b,\label{buoyalign}
    \\ \frac{\partial \Omega}{\partial t}&=-u'\frac{\partial \Omega}{\partial x'}-w'\frac{\partial \Omega}{\partial z'}-\Ri \cos \phi \frac{\partial b}{\partial x'}+\Ri \sin \phi \frac{\partial b}{\partial z'}+\frac{1}{\Reynolds}\nabla'^2\Omega,\label{voralign}
\end{align}
where $\Omega=\partial u'/\partial z'-\partial w'/\partial x'$ is the vorticity, and $u'$ and $w'$ refer to the braid-aligned velocities. In this long and thin braid flow, we assume gradients in $z'$ are greater than those in $x'$, thus neglecting the $\partial/\partial x'$ terms and identify vorticity as nearly equal to the cross-braid shear. The baroclinic amplification of shear is due to the term $\Ri \sin \phi \, (\partial b/\partial z')$ in \eqref{voralign}. We further assume a uniform compressive strain field $w' = -\gamma z'$ local to the braid (to be validated later by simulations), where $\gamma(t)$ is the local strain rate. These assumptions are particularly accurate far away from the cores and reduce \eqref{buoyalign} and \eqref{voralign} to 
\begin{equation}\label{alignlins}
\refstepcounter{equation}
\begin{aligned}
    \frac{\partial b}{\partial t} 
    &\approx \gamma z'\frac{\partial b}{\partial z'} 
    + \frac{1}{\Reynolds\,\Prandtl}\frac{\partial^2 b}{\partial z'^2}, 
    \qquad
    \frac{\partial \Omega}{\partial t}
    \approx \gamma z'\frac{\partial \Omega}{\partial z'}
    + \Ri \sin \phi \frac{\partial b}{\partial z'}
    + \frac{1}{\Reynolds}\frac{\partial^2 \Omega}{\partial z'^2}.
\end{aligned}
\tag{\theequation a--b}
\end{equation}

We seek similarity solutions
\begin{equation}
    b(z,t) = F(\xi_b),\quad \Omega(z,t) = S(t) \, G(\xi_\Omega),\label{eq:F,G}
\end{equation}
where $\xi_b = z'/\delta_b(t)$ and $\xi_\Omega=z'/\delta_\Omega(t)$ for time-dependent buoyancy and velocity layer thicknesses $\delta_b$ and $\delta_\Omega$.  We further consider the inviscid limit $\Reynolds \to \infty$ in \eqref{alignlins}, expected to hold as long as $\delta_b$ and $\delta_\Omega$ are sufficiently large compared to the length scales at which diffusion of buoyancy and velocity are $O(1)$ (see Supplementary Material, section 1 for details\footnote{Supplementary Material is available as an ancillary file on arXiv.}). The above similarity solutions satisfy \eqref{alignlins} provided that (i) $\Reynolds \to \infty$; (ii) $G=\partial F/\partial \xi$, where $\xi\equiv \xi_b = \xi_\Omega$ or equivalently $\delta(t)\equiv \delta_b(t)=\delta_\Omega(t)$, valid with initial conditions \eqref{ICnd}; (iii) and that $\delta(t)$ and the baroclinically generated shear $S(t)$ satisfy
\begin{equation} \label{ddeltadt_dSdt}
\refstepcounter{equation}
\begin{aligned}
    \frac{\partial \delta}{\partial t}=-\gamma \delta, \qquad \frac{\partial S}{\partial t}= \frac{\Ri \sin\phi}{\delta}.
\end{aligned}
\tag{\theequation a--b}
\end{equation}
Note that \eqref{eq:F,G} implies $\partial b/\partial z|_{z=0} = \delta^{-1} \partial F/\partial \xi(0)$, which, with the given initial conditions, corresponds to $\delta(0)\approx 1$ and $\partial F/\partial \xi(0)\approx 1$.
With the far-field boundary conditions $F(\pm\infty)=\pm 1$, CS76 showed that retaining the diffusive terms in \eqref{alignlins} determines $F(\xi)$ explicitly in the form of the solution to a 1D diffusion equation. In the inviscid limit however, the initial conditions $F(\xi) = b(z,t=0)$ are never `forgotten' and the CS76 solution is invalid, except in the particular case when the initial velocity and buoyancy profiles take the form of error functions (as they assumed).

For simplicity, here we assume $\gamma(t)=\gamma_1t+\gamma_0$ and $\phi(t)=\phi_1t+\phi_0$ during the initial period of billow roll up (to be validated later by simulations), where $\gamma_0$, $\gamma_1$, $\phi_0$ and $\phi_1$ are anticipated to be empirical functions of $\Ri$ and of the initial conditions. This approach may be compared with CS76, who derived an inviscid model for the advection of vorticity from the braid into the cores that was then solved numerically to compute $\gamma(t)$ and $\phi(t)$.  Note that, for comparison with our  simulations, the initial angle $\phi_0$ and strain $\gamma_0$ are both small but non-zero. The evolution of braid thickness $\delta(t)$ is then given by
\begin{align}
    \delta(t) = \exp\left[-\int_0^t \gamma(\tau) \;d\tau\right] = \exp\left[- \frac 1 2 \gamma_1 t^2 -\gamma_0 t\right],\label{deltaeq}
\end{align}
whilst the evolution of braid shear $S(t)$ is given by
 \begin{equation} \label{S'eq}
 S(t) = S(0) + \int_0^{t}\Ri \,\sin\left(\phi_1 \tau + \phi_0 \right) \,\exp\left[\frac 1 2 \gamma_1 \tau^2 +\gamma_0 \tau\right] \! \diff \tau .
 \end{equation}

\subsection{Braid stability criterion}
CS76 viewed the braid as a steady parallel stratified shear layer, where a necessary criterion for instability is that the minimum gradient Richardson number at the centre  is
\begin{equation}
    \Rig(0,t) = \Ri \frac{ (\partial b/\partial z) \cos\phi}{\Omega^2} \Big|_{z'=0} = \frac{\Ri\cos\phi }{\delta(t) S(t)^2}<1/4.\label{eq:Rig}
\end{equation}
They argued that, under this condition, SSI may develop, assuming that the timescale of the local braid flow (which dictates the roll-up time of subsequent shear instabilities) is much shorter than that of the background shear flow. In our notation, this corresponds to $S\gg 1$. However, identifying a threshold value is difficult as the assumptions underlying  standard linear stability analysis  break down in the braid. In particular, the strain field acts to damp growing perturbations. 

To address this limitation, \citet{staquet_two-dimensional_1995} used results from the study of a strip of vorticity under uniform straining by \citet{dritschel_stability_1991} to argue that a stricter criterion for SSI onset should involve the ratio of strain to vorticity (shear):
\begin{equation}\label{eq:strain_criterion}
    \frac{\gamma}{\Omega}\Big|_{z'=0} = \frac{\gamma_1t +\gamma_0}{S(t)} < \Gamma^*.
\end{equation}
In this picture, SSI is in fact the product of rapid transient (non-normal) growth that can reach significant amplitude before the strain is able to damp perturbations. Critical values of $\Gamma^*$ between $0.02$ and $0.1$ have been suggested \citep{staquet_two-dimensional_1995,  smyth_secondary_2003,mashayek_zoo_2012-1}. Importantly, however, these studies assumed that SSI develops at a time when braid thickness $\delta$ has reached an equilibrium in which straining is balanced by viscous diffusion, as was verified at the time using simulations at $\Reynolds < 10^4$. This viscous equilibrium occurs approximately at time $t_{2D}$, defined as the time at which the amplitude of the primary billow saturates and the braid strain becomes approximately constant.

Considering instead the earlier period of braid development, a small angle approximation in \eqref{S'eq} together with the observation that $\phi_0$ and $\gamma_0$ are small yields  $S(t)\sim \exp(\gamma_1 t^2/2)$ for large $t$. This implies that the timescale over which the threshold  \eqref{eq:strain_criterion} is reached (i.e., when the baroclinic shear becomes dominant) is expected to scale approximately as $1/\sqrt{\gamma_1}$. It will be shown below that this timescale may be considerably shorter than that of primary billow saturation $t_{2D}$, allowing for what we term early onset of SSI.





\section{Numerical simulations}\label{numsim}
\begin{figure}
    \centering
    \includegraphics[width=\linewidth]{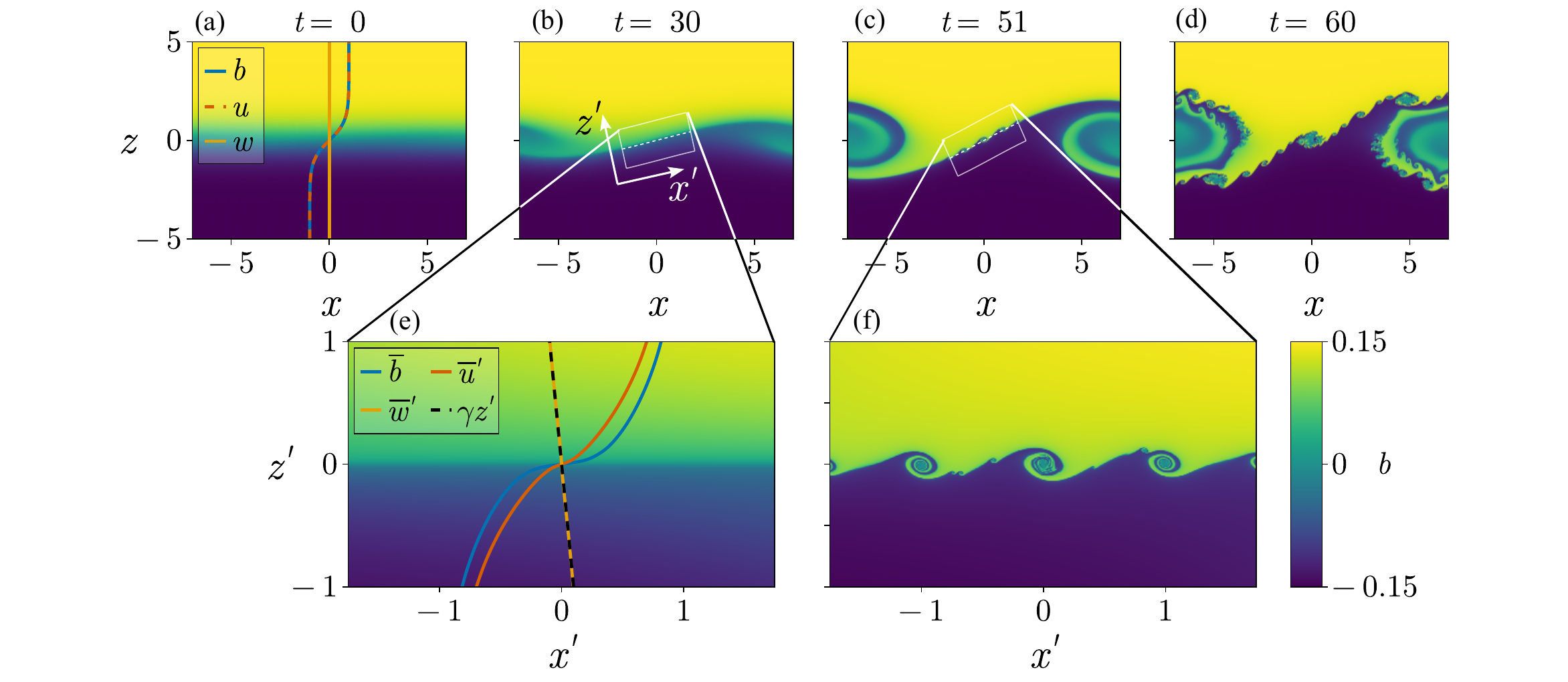}
    \caption{ An example of the numerical braid analysis for a simulation with $(\Ri,\Reynolds)=(0.15,10^6)$, showing (a-d) the transient evolution of the buoyancy field $b$. The initial profiles for $b$, $u$ and $w$ are shown on panel (a). The white dashed lines in panels (b) and (c) illustrate the location of the braid, and a white box shows the local braid field. The local buoyancy field $b(x',z')$ is shown in (e) and (f). Panel (e) also overlays the $x'$-averaged profiles $\overline b(z')$, $\overline u'(z')$ and $\overline w'(z')$. The black dashed line in panel (e) shows the linear approximation for $w'$ used to diagnose the strain $\gamma(t)$.}\label{basic_sim}
\end{figure}
\subsection{Setup}
We run direct numerical simulations (DNS) using Oceananigans.jl, a GPU-accelerated finite volume solver \citep{wagner_high-level_2025} with a centred second-order advection scheme and constant isotropic diffusivities. We present a set of 16 simulations with geophysically-relevant $\Ri = [0.05, 0.10, 0.15, 0.20]$ and $\Reynolds = [10^4,10^5, 10^6, 10^7]$. Simulations are limited to $\Reynolds\ge 10^4$ as lower values do not exhibit early SSI onset (see Supplementary Material, section 1).

The flow is horizontally periodic in the $x$ domain $[-L_x/2,\; L_x/2]$ and is subject to  free-slip ($\partial u/\partial z=0$) and no-flux  ($w=0$ and $\partial b/\partial z=0$) boundary conditions at $z =-L_z/2,\;L_z/2 $. The length $L_x$ is chosen to be the wavelength of the fastest growing linear mode for each $(\Ri, \Reynolds)$  --- typically $L_x\approx 14$ --- thus restricting the domain to one billow length and excluding the pairing instability \citep{smyth_secondary_2003}. The  domain height $L_z=20$ is chosen sufficiently large to minimise boundary effects on instability growth.  A fixed stretched grid was used in $z$ to increase resolution in the central region where gradients are sharpest, with grid spacing at $z=0$ five times finer than the average. To initiate KH instability growth, the flow is perturbed with the fastest growing linear mode corresponding to each $(\Reynolds, \Ri)$ with  energy $10^{-4}$ of the base flow. All cases were run on a grid of $8000\times 8000$ points, which was sufficient to ensure convergence (more details in  the Supplementary Material, section 2).

\Cref{basic_sim} shows the evolution of a sample simulation, illustrating the initial condition (\cref{basic_sim}a), primary billow growth (\cref{basic_sim}b) and subsequent onset of SSI on the braid (\cref{basic_sim}c). \Cref{basic_sim}d shows the subsequent advection of secondary instabilities towards the cores and the ultimate breakdown of the primary and secondary billows. These dynamics are beyond the scope of this study, as they will clearly be dominated by three-dimensional instabilities \citep{mashayek_zoo_2012-1,mashayek_zoo_2012}. Therefore, simulations are considered up to the time of primary billow saturation $t_{2D}$ or the time of SSI onset $t_S$, whichever occurs first --- the valid 2D regime. Note that $t_S$ is defined as the time at which $\Rig$ reaches a minimum, as SSI growth causes a subsequent increase.

\subsection{Numerical braid diagnostics}\label{braidanalysis}
To allow for comparison between the predictions of the inviscid braid evolution model, the simulation outputs were analysed as follows. First, the initial linear mode perturbation was positioned so that the braid inflection point is always located at the origin $(x,z)=(0,0)$. Second, we defined the braid following the fitting procedure outlined in \cite{smyth_secondary_2003}, fitting a cubic polynomial to the points of strongest shear, as seen by white dashed lines in figures \ref{basic_sim}b-c. 
Third, we calculate the average braid angle $\phi$ from the gradient of the braid locus. Fourth, transects taken along the braid inclined at angle $\phi$ reconstructed the buoyancy $b$, along-braid shearing velocity $u'$ and cross-braid straining velocity $w'$. Figures \ref{basic_sim}e and f illustrate $b$ where the braid is now centred around $z'=0$.  Before the growth of secondary instabilities, the flow is approximately parallel along $x'$, and so we average along $x'$ to construct averaged profiles of buoyancy $\overline b(z')$, and velocities $\overline u'(z')$ and $\overline w'(z')$; such profiles are overlaid in \cref{basic_sim}e. The braid is similar to the initial base flow in \cref{basic_sim}a, but is subject to additional straining. These averaged profiles are used to calculate $\Rig$  and $\gamma/\Omega$, as defined in \eqref{eq:Rig} and \eqref{eq:strain_criterion}, respectively. Finally, a linear fit $\overline w'=-\gamma z'$ (\cref{basic_sim}e) validates our earlier assumption of uniform strain and provides the time-evolving strain $\gamma(t)$ to be used in the inviscid model. 

\section{Results}\label{results}
\begin{figure}
    \centering
    \includegraphics[width=\linewidth]{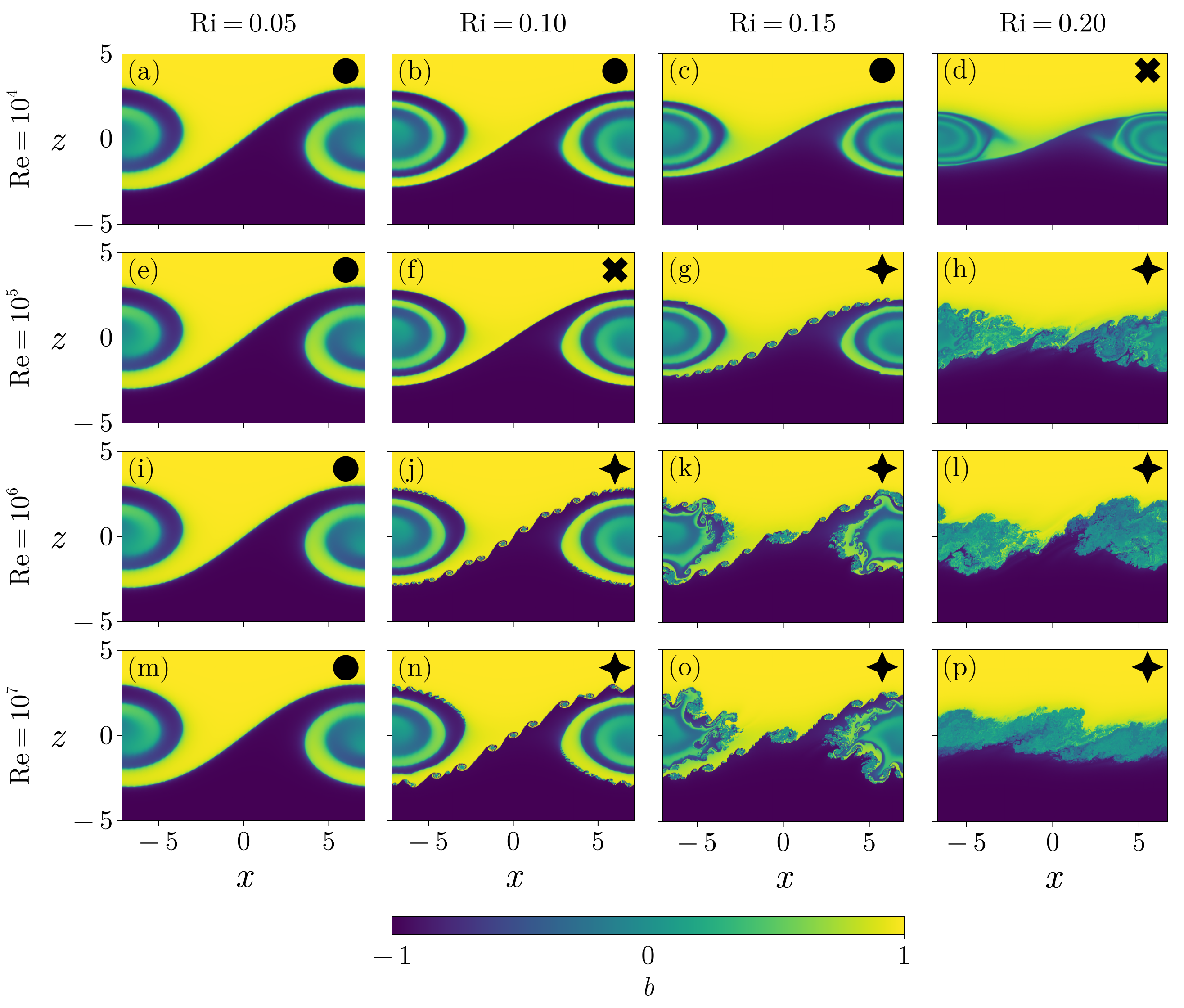}
    \caption{Snapshots of the buoyancy field $b$ at the time of primary billow saturation, $t_{2D}$, over a range of $\Reynolds$ (rows) and $\Ri$ (columns). Symbols in the upper right corners represent the stability outcome of the braid before $t_{2D}$; stars denote SSI, crosses denote marginal SSI (instigated by additional noise, not shown here), and circles denote absence of SSI. }
    \label{t2D}
\end{figure}

\subsection{Overview and SSI onset in $(\Ri,\Reynolds)$ space}
\Cref{t2D} presents an overview of the full parameter range using snapshots of the buoyancy field at $t_{2D}$. Qualitatively, \cref{t2D} confirms the well-known result that as $\Ri$ increases (from left to right), the amplitude of the primary billows decreases. As stratification increases with $\Ri$, greater work is required to raise the denser fluid in the billow, resulting in flatter and weaker cores. The role of increasing $\Reynolds$ (from top to bottom) is less apparent but clearly $\Ri$-dependent; at $\Ri=0.05$, the billows appear identical across all $\Reynolds$ since there is no onset of SSI. However, above this $\Ri$, increasing $\Reynolds$ favours SSI. At higher $\Ri$, SSI occurs at a lower $\Reynolds$. While \cref{t2D} shows that high $\Reynolds$ and $\Ri$ are generally favourable to the onset of SSI before primary billow saturation ($t_S<t_{2D}$), a more subtle classification is needed.

\begin{figure}
    \centering
    \begin{subfigure}[c]{\textwidth}
        \centering
        \includegraphics[width=0.95\linewidth]{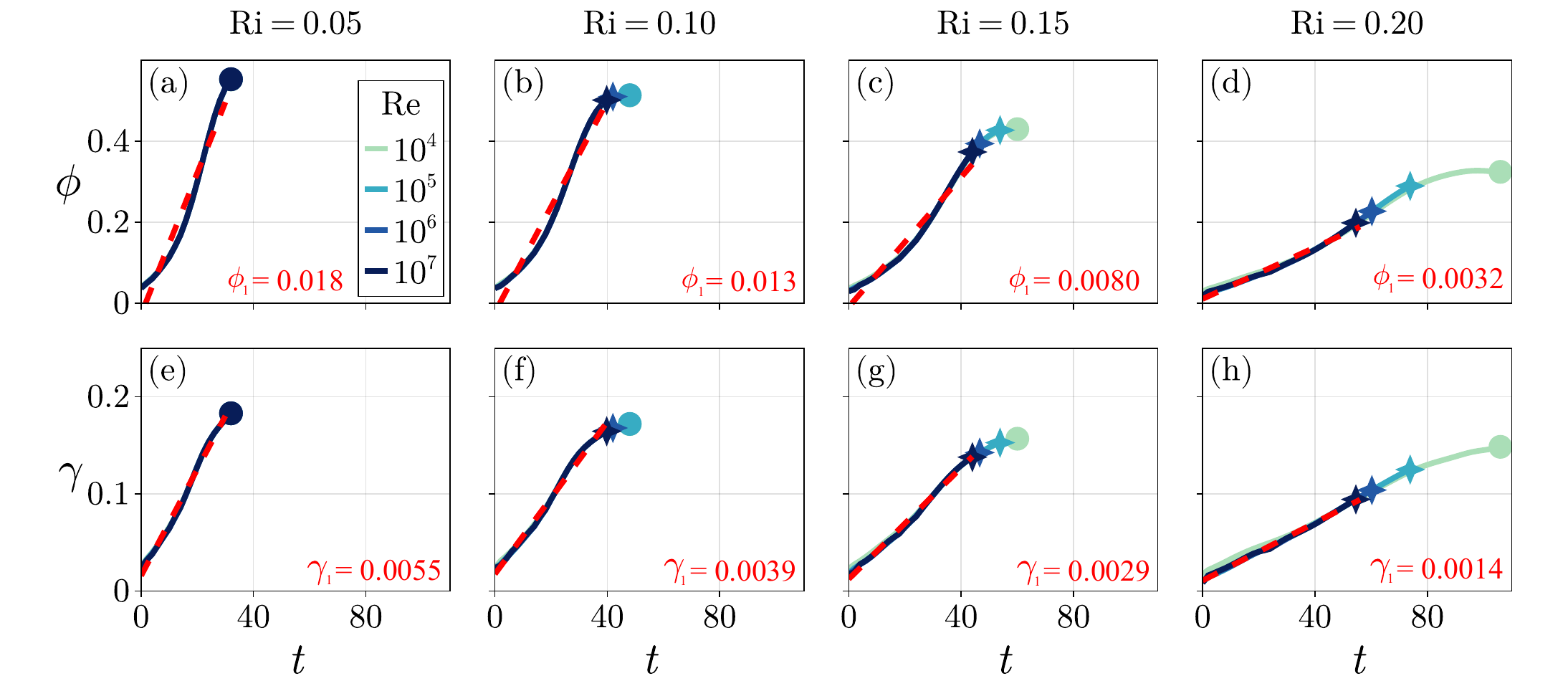}
    \end{subfigure}
    \caption{The temporal evolution of braid angle $\phi$ and strain $\gamma$ at $\Reynolds=[10^4,10^5,10^6,10^7]$, as given by the legend, and $\Ri=[0.05,0.10,0.15,0.20]$, as given by column titles. The primary billow saturation is marked at $t_{2D}$ by a circle (see corresponding snapshots in figure~\ref{t2D}), and SSI onset is marked at $t_S$ by a star. Red dashed lines and slopes $\phi_1,\gamma_1$ show the linear fits to be used in the inviscid model for $\gamma/\Omega$ (see \cref{{gamS}}(e-h)).}
    \label{Primary1}
\end{figure}

The transient billow growth is quantified in \cref{Primary1}, where the growth rate and maximal values of both $\phi$ (a proxy for core amplitude) and $\gamma$ (a proxy for core rotation rate) decrease with $\Ri$. This billow growth is $\Ri$-dependent and essentially $\Reynolds$-independent. When SSI occurs, higher $\Reynolds$ results in an earlier SSI onset  (smaller $t_S$). The red dashed lines in \cref{Primary1} represent the linear fits $\phi= \phi_1 t+ \phi_0$ and $\gamma=\gamma_1 t +\gamma_0$ used as inputs in the inviscid model. The slopes $\phi_1$ and $\gamma_1$ (given in red) are found to decrease approximately linearly with $\Ri$.

\subsection{Braid evolution and stability}\label{sec:braidstab}
\begin{figure}
    \centering
    \begin{subfigure}[c]{\textwidth}
        \includegraphics[width=0.95\linewidth]{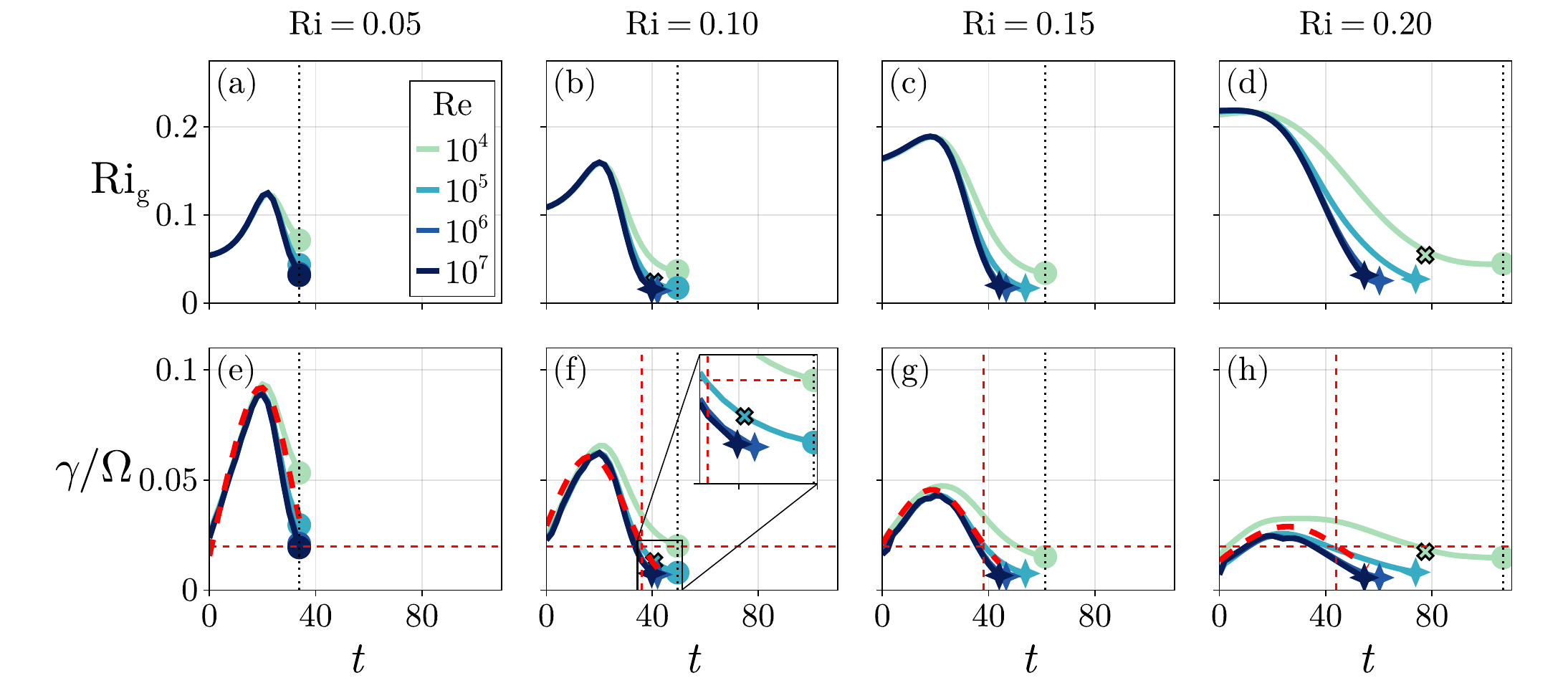}
    \end{subfigure}
    \caption{The temporal evolution of the gradient Richardson number $\Rig$ and ratio of strain to shear $\gamma/\Omega$. Symbols (circles, stars and crosses) indicate the braid stability outcome as in \cref{t2D}. The red dashed curve in the lower row is the inviscid model prediction. Horizontal red dashed lines show the critical $\Gamma^* =0.02$, while vertical red dashed lines indicate the time $t^*$ at which the inviscid model attains this critical value. Vertical black dotted lines indicate the time $t_{2D}$. Note that the separation $t_{2D}-t^*$ increases with $\Ri$ for $\Ri\ge 0.10$.}
    \label{gamS}
\end{figure}
The evolution of the minimum gradient Richardson number $\Rig(0,t)$ is shown in \cref{gamS} (top row). Early on, the braid is not dynamically distinct from the primary billow and $\Rig$ is effectively independent of $\Reynolds$. During this stage, straining thins the braid, increasing local stratification and  raising $\Rig$, effectively stabilising the flow. However, once shear begins to intensify, this trend reverses and $\Rig$  decreases rapidly, at which point differences between $\Reynolds$ cases become apparent.


While $\Rig$ is informative, it is insufficient to diagnose SSI onset as it remains below $1/4$ throughout, and there is no other apparent useful critical value. All simulations attain similarly low values $\Rig\lesssim 0.05$ before $t_{2D}$, emphasising the weakly stratified nature of the braid. This motivates the comparison to the stability of a strained strip of vorticity and the choice of a stability criterion based on $\gamma/\Omega$, whose evolution is shown in the bottom row of \cref{gamS}. Again, all growth is initially $\Reynolds$-independent; the initial increase of $\gamma/\Omega$ is approximately linear and governed solely by the increasing strain from the primary billow. As baroclinic acceleration then dominates, $\gamma/\Omega$ decreases. The onset of SSI occurs below a critical $\Gamma^*=0.02$, which emerges as a suitable threshold across the parameter space studied here. We define $t^*$ as the time at which this threshold is first reached in the inviscid model. The onset of SSI is limited by the available time window $t_{2D}-t^*$ over which transient growth can amplify. This window increases with $\Ri$, as shown in \cref{gamS} (lower row), as $t_{2D}$ grows more rapidly with $\Ri$ than $t^*$ (details in Supplementary Material, Section 3).

To further confirm this interpretation, all simulations that reached $\Gamma^*$ but did not exhibit SSI were restarted $\approx 10$ time units before $t^*$ with the addition of divergence-free white noise with modest energy $10^{-4}$ of the initial flow. This was sufficient to trigger SSI in two marginal cases, marked with crosses on \cref{t2D} and \cref{gamS}: $(\Ri,\Reynolds)=(0.10,10^5)$ and $(0.20,10^4)$. By contrast, this additional noise did not trigger visible SSI prior to $t_{2D}$ at the lower values of $(\Ri, \Reynolds) = (0.10,10^4)$ and $(0.15,10^4)$, where transient growth did not reach finite amplitude.

 \subsection{Agreement with the inviscid model at finite $\Reynolds$}

Finally, \cref{gamS} compares the simulations at increasing $\Reynolds$ with the inviscid model predictions for $\gamma/\Omega$ (red dashed curves in the bottom row) from \eqref{S'eq}, \eqref{eq:strain_criterion} and the linear fits for $\phi,\gamma$ (exact values in Supplementary Material, Section 3). As expected, the agreement improves with increasing $\Reynolds$. The inviscid predictions for $t^*$  (vertical red dashed line) are accurate for $\Reynolds\ge 10^5$. The lower $\Reynolds=10^4$ cases deviate from the inviscid evolution only after the braid has thinned enough for viscous diffusion to act and limit further thinning, resulting in weaker baroclinic shear amplification than predicted. In cases with intermediate $\Ri$, for which $t^*\lesssim t_{2D}$  already in the inviscid limit, even relatively modest viscous effects are sufficient to prevent the early onset of SSI altogether, as is observed for $\Ri = 0.10$ and $0.15$ at $\Reynolds=10^4$. 

\section{Conclusions}\label{discussion}
We studied the mechanism leading to the early onset of SSI on the braids between KH billows over an extended range of Richardson and Reynolds numbers overlapping with ocean observations that frequently show braid-dominated turbulent mixing. To do so, we constructed a semi-analytical inviscid model (see \eqref{deltaeq}, \eqref{S'eq}) for the transient evolution of the braid, comparing the results to high-resolution two-dimensional DNS. 

As the braid steepens and thins under the influence of the growing primary billow, its stability reflects a competition between  stabilising strain, which decreases with $\Ri$ (due to flatter billows), and destabilising  baroclinicity, which amplifies the shear and increases with $\Ri$ (see \eqref{ddeltadt_dSdt}). This balance is captured by the ratio of strain to braid-aligned shear, $\gamma/\Omega$ (see \eqref{eq:strain_criterion}). Using linear approximations for $\gamma$ and the braid angle $\phi$ during primary billow growth, the timescale over which the destabilising baroclinic shear becomes dominant is approximately given by $1/\sqrt{d\gamma/dt}$, highlighting the importance of the time-evolving braid flow.
More precisely, numerical simulations confirm that SSI can first grow at time $t^*$ when $\gamma/\Omega< \Gamma^* = 0.02$. In the inviscid limit, early SSI requires $\Ri \gtrsim 0.10$,  for which sufficient separation develops between $t^*$ 
and the primary billow saturation time $t_{2D}$ as $\Ri$ increases, providing sufficient time for SSI to grow to finite amplitude.

As summarised in \cref{t2D}, simulations confirm early SSI onset for $\Ri \gtrsim 0.10$  at $\Reynolds=10^6$ and $10^7$, where the braid dynamics remain essentially inviscid and well predicted by our model.  At $\Reynolds=10^5$, viscosity slightly slows braid thinning, delaying $t^*$ and narrowing the window $t_{2D}-t^*$ available for SSI growth ($t_{2D}$ being independent of $\Reynolds$), rendering $\Ri = 0.10$ marginal, with onset requiring additional noise (likely present in natural flows). At $\Reynolds=10^4$, viscosity further slows and nearly halts braid thinning, thus preventing SSI at $\Ri = 0.10$ and $0.15$, and leaving only a marginal case at $\Ri=0.20$.



We note that our simulations are two-dimensional and permit only a single primary billow wavelength, thus precluding vortex pairing and convective instabilities. Nonetheless, when SSI outpaces primary billow saturation, it should precede both. The growth rate of vortex pairing decreases rapidly with $\Ri$ \citep{klaassen_role_1989}, and is sufficiently small at $\Ri = 0.10$ that pairing generally occurs well after $t_{2D}$. Similarly, \citet{mashayek_zoo_2012-1,mashayek_zoo_2012} show that convective instabilities become significant only near or after $t_{2D}$. As noted in the introduction, other types of braid instabilities are possible beyond SSI. Although SSI is the first to emerge in all of our simulations, dynamics resembling those of the braid stagnation point instability described by \citet{mashayek_zoo_2012-1} are observed by $t_{2D}$ in some cases (e.g. panels \textit{k} and \textit{o} of \cref{t2D}).

An avenue for future work is to investigate the turbulence generated by the early onset of SSI, and its impact on the subsequent evolution of KH billows in three-dimensional DNS. As first noted by CS76, rapid mixing in the braid may strongly affect billow dynamics, and field observations at $\Reynolds = O(10^5-10^7)$ indeed show intense braid mixing and weak mixing in billow cores, suggesting that braid turbulence may arrest primary billow growth.
However, even in the absence of fully developed turbulence, rapid  braid thinning during primary KH billow growth poses a significant computational challenge at high $\Reynolds$. Poorly resolved interfaces can trigger spurious instabilities that vanish with increasing resolution \citep{lewin2021influence,lecoanet_validated_2016}. The simple framework introduced here to estimate the interface thickness $\delta$ provides a criterion for the minimum grid spacing required to capture SSI onset. In our simulations, $\delta > 10\Delta z $ at the time of SSI onset. In three-dimensional DNS, this constraint should be considered alongside, and may be more restrictive than, resolution criteria based on turbulent Kolmogorov or Batchelor scales.

Finally, oceanic flows have $\Pr = O(10\text{–}10^3)$ --- well above the $\Pr = 1$ considered here. At geophysically relevant $\Reynolds \ge 10^6$, we showed that the braid dynamics are effectively inviscid, so our results for SSI onset across $(\Ri,\Reynolds)$ space should  be independent of $\Pr$. At lower $\Reynolds = O(10^4-10^5)$, however, $\Pr\gg 1$ can influence braid evolution and SSI onset, particularly near marginal $(\Ri,\Reynolds)$ values. A higher $\Pr$ will reduce the interface thickness at which diffusion hinders strain-induced thinning, thus promoting closer agreement with the inviscid model and favouring SSI. However, supplementary simulations (Supplementary Material, section 4) show that it also induces non-trivial changes in the temporal evolution of $\Rig$ and $\gamma/\Omega$. These high-$\Pr$ effects likely arise from the decoupling of buoyancy and shear interface thicknesses, which is not captured by our model and remains to be investigated.
\backsection[Funding statement]{The authors acknowledge the support of the Geophysical Fluid Dynamics Program at the Woods Hole Oceanographic Institution, supported by the NSF and ONR. E. R. Bouckley was supported by Saint Gobain. A. Lefauve was supported by a NERC Independent Research Fellowship NE/W008971/1.

This work used the Delta advanced computing and data resource through allocation PHY250188 from the Advanced Cyberinfrastructure Coordination Ecosystem: Services \& Support (ACCESS) program, which is supported by NSF grants \#2138259, \#2138286, \#2138307, \#2137603, and \#2138296. Delta is a joint effort of the University of Illinois Urbana-Champaign and its National Center for Supercomputing Applications, supported by the NSF (award OAC 2005572) and the State of Illinois.
 }
 \backsection[Acknowledgments]{We thank A. Kaminski for helpful discussions and A. Atoufi and J. R. Taylor for sharing code that was adapted for the simulations.}
\backsection[Declaration of interests]{The authors report no conflict of interest.}
\bibliographystyle{jfm}
\bibliography{references,AL_references}

\end{document}